\documentclass[conference]{IEEEtran}

\usepackage{algpseudocode}
\usepackage[ruled,vlined,linesnumbered,noend]{algorithm2e}
\usepackage{xcolor}
\usepackage{graphicx}
\usepackage{textcomp}
\usepackage{enumitem}
\usepackage{courier}
\usepackage{wrapfig,lipsum}

\usepackage[most]{tcolorbox}

\usepackage{xcolor,colortbl}
\usepackage{booktabs}
\usepackage{multirow}
\usepackage{pifont}
\usepackage{bbding}
\usepackage{soul}
\usepackage{url}                
\usepackage{xcolor}             
\usepackage[]{hyperref}         
\hypersetup{                    
  colorlinks,
  linkcolor={green!80!black},
  citecolor={red!70!black},
  urlcolor={blue!70!black}
}
\usepackage{url}
\usepackage{xcolor}
\usepackage{tcolorbox}
\newcommand{\tech}{\mbox{\textsc{PoVGen}}}

\usepackage{breakurl}

\usepackage{tikz}
\usetikzlibrary{arrows.meta,positioning,fit,calc,backgrounds,shapes.geometric}

\begin{document}
%
%
%


\title{Neuro-Symbolic Proof-of-Vulnerability Generation with Open-Weight Models}

\author{
\IEEEauthorblockN{Yu Nong}
\IEEEauthorblockA{\textit{University at Buffalo}\\
Buffalo, NY, USA \\
yunong@buffalo.edu}
\and
\IEEEauthorblockN{Haipeng Cai}
\IEEEauthorblockA{\textit{University at Buffalo}\\
Buffalo, NY, USA \\
haipengc@buffalo.edu}
}

\maketitle


%

\thispagestyle{plain}
\pagestyle{plain}
\pagenumbering{arabic}

\begin{abstract}

Software vulnerabilities are persistent, but validating 
them remains difficult: a \emph{Proof-of-Vulnerability (PoV)} requires a concrete input that triggers the vulnerable behavior, yet public triggering inputs are often unavailable for disclosed vulnerabilities. 
%
Existing techniques make different tradeoffs in effectiveness, scalability, cost, and controllability, leaving room for complementary designs.
%
To complement them, we present {\tech}, a low-cost neuro-symbolic framework that makes PoV generation cost-effective via \emph{semantic focusing} and 
\textit{LLM-guided constraint reasoning} using \textit{open-weight} models. 
{\tech} first localizes vulnerability-relevant regions (utilizing patch information if available), then performs path-sensitive reachability analysis, and finally generates PoVs by extracting and solving constraints with LLM-guided reasoning backed by an SMT solver. 
%
%
{\tech} successfully generates PoVs for 78.98\% of vulnerabilities in a recent benchmark, outperforming fuzzing (up to 50.20\%) and symbolic execution (2.45\%). 
On 250 real-world CVEs without public PoVs, it generates valid PoVs for 74.80\% of cases and reproduces 65.1\% when without patch information. 
The fine-tuned open-weight models match frontier commercial LLMs on key sub-tasks (i.e., the core constraint-reasoning steps) while running locally at no per-sample API cost. Applying the generated PoVs revealed six flawed patches in disclosed CVEs (all subsequently fixed) and five previously unreported vulnerabilities (of which four have been confirmed and fixed by the developers).

\end{abstract}

\vspace{-2pt}
\section{Introduction}\label{sec:intro}
\vspace{-2pt}

Software vulnerabilities are prevalent~\cite{cvedashboard23} and consequential~\cite{vulconsequence231}, making timely and effective vulnerability patching key to cyber defense~\cite{pearce2023examining,zhou2024largeb,wu2023effective}.
However, a patch alone often leaves a critical evidentiary gap: it indicates that developers modified security-relevant code, but does not by itself provide a concrete way to reproduce the vulnerable behavior or validate the effectiveness of the fix. 

A \emph{Proof-of-Vulnerability (PoV)}—an input that triggers the vulnerability in a program~\cite{mei2024arvo}—fills this gap by turning a disclosed vulnerability into a concrete, reproducible security condition.
A PoV enables vendors to validate reported issues, allows defenders to test mitigations, and provides researchers with unambiguous evidence of vulnerability impact. It is also essential for verifying patches: prior studies show that 
patches can be incomplete and regressive~\cite{li2017large}; a working PoV helps defenders confirm whether a patch actually closes the vulnerability~\cite{wu2025veribin,kikta2024patch}. 
Yet in practice PoVs remain scarce. 


Efforts to address this gap exist but remain limited. Datasets such as ARVO~\cite{mei2024arvo} curate reproducible vulnerabilities from systems like OSS-Fuzz~\cite{serebryany2017oss}, but cannot generate PoVs for the broader set of disclosed CVEs beyond those already accompanied by reproducible triggering inputs. 
Fuzzing is often ineffective for triggering specific, known vulnerabilities due to its broad search space~\cite{stephens2017driller}, while symbolic execution~\cite{cadar2008klee} struggles to scale to large codebases and complex execution environments. Other approaches target narrower classes of vulnerabilities (e.g., memory corruption~\cite{wang2025pbaeg}, object injection~\cite{park2022fugio,marques2025explode}) or specific execution environments (e.g., kernels~\cite{chen2020koobe}, web applications~\cite{alhuzali2018navex}).


Recently, 
LLM-assisted systems emerged as promising solutions. 
For instance, LLMs used in a synthesis-and-validate loop can help exploit vulnerable functions 
in Npm packages~\cite{simsek2025pocgen}. The reliance on the model to implicitly handle path and constraint reasoning can limit scalability to real-world CVEs. 
LLMs have also empowered concolic execution~\cite{tu2026cottontail,luo2026agentic}, which can generate more test inputs 
from dynamic runs on existing seeds, or even further generating seeds with LLMs~\cite{tu2026cottontail}. 
These approaches rely on closed-weight, often costly LLMs fully under vendor's control, and their objective is coverage, not triggering a specific vulnerability.
Other work~\cite{li2025large} shows that open-weight LLMs are also capable of symbolic execution, yet designed for verifying post-conditions. 


In this paper, we explore a complementary direction: low-cost, vulnerability-specific PoV generation that combines symbolic path guidance with fine-tuned \textit{open-weight} models to generate concrete triggering inputs while reducing reliance on proprietary model inference.
%
%
We view this setting as a \emph{vulnerability-targeted input synthesis problem}: the system must identify execution paths that reach vulnerable behavior and synthesize concrete inputs that drive the program along those paths. Yet, this setting raises three challenges. \textit{First}, the lack of a vulnerability-triggering input leaves a large and weakly structured search space, making unguided exploration inefficient. 
\textit{Second}, vulnerability-triggering constraints are difficult to recover in practice as they often depend on program state, pointer relationships, input format, and external environment/library behaviors.
\textit{Third}, LLM-assisted reasoning can be expensive if it repeatedly invokes frontier proprietary models over many candidate paths, and such reliance also reduces control over model behavior, availability, and cost.


These challenges impose three key requirements: (1) narrowing the search space to vulnerability-relevant regions before path exploration, (2) combining symbolic reasoning with model-guided constraint extraction and input synthesis, and (3) keeping model dependence and inference cost under control. We address these requirements with {\tech}, a low-cost framework for PoV generation using \textit{open-weight} models. 
This openness makes the system more \textit{controllable} (i.e., allowing the model artifact itself to be pinned, audited, and reused), 
which matters for security tooling as model changes can affect both output behavior and experimental repeatability. 


{\tech} works in two modes: a \texttt{patch-guided} mode that uses patch information as a guide, and a \texttt{patch-free} mode for scenarios where patches are unavailable (e.g., unpatched or newly disclosed vulnerabilities). It decomposes PoV generation into three stages. First, it uses a fine-tuned open-weight model to identify the vulnerability manifestation point, restricting analysis to vulnerability-relevant regions. Second, it performs guided exploration 
to identify interprocedural paths from program entry points to the manifestation point, reducing a large codebase to a small set of candidate paths. Third, it extracts and solves path constraints along these paths using LLM-guided reasoning backed by an SMT solver, using two dedicated fine-tuned open-weight models. 

Our evaluation demonstrates that {\tech} is both effective and cost-efficient. On the ARVO benchmark, {\tech} generates valid PoVs for 78.98\% of vulnerabilities in the \texttt{patch-guided} mode, outperforming fuzzing baselines (up to 50.20\% success) and symbolic execution (2.45\%). In controlled comparisons, {\tech}'s task-specialized open-weight models also show stronger task-level performance than direct frontier-LLM prompting while substantially reducing model cost. For example, the fine-tuned constraint-extraction model achieves 89.63\% semantic recall, compared to 57.04\% by the same base open-weight model without fine-tuning and 86.14\% by Claude-Sonnet-4 (at \$2.06 per sample, versus no per-sample API cost for {\tech}'s locally deployed model). 
On 250 real-world CVEs without public PoVs, {\tech} generates valid PoVs for 74.80\% of the vulnerabilities, and achieves 65.1\% success in the \texttt{patch-free} mode.


Beyond validating known vulnerabilities, the generated PoVs also enabled patch validation and vulnerability discovery: we identified six flawed patches in disclosed CVEs, all of which have been confirmed and fixed in later versions, and discovered five previously unreported vulnerabilities, four of which have been confirmed and fixed by upstream maintainers, with one CVE ID assigned so far. {\tech}'s modular design further supports extension to other vulnerability types and programming languages, and its components can be used independently for other tasks such as constraint-guided input synthesis and vulnerability localization.

In summary, this work makes the following contributions:

\begin{itemize}[leftmargin=*,itemsep=0pt,topsep=2pt]
\item 
We characterize PoV generation 
as a vulnerability-targeted input synthesis problem, and identify the need for search-space focusing, constraint-guided input synthesis, and cost-controlled model reasoning ($\S$\ref{sec:bkgmot}).

\item 
We design {\tech}, a low-cost PoV-generation framework that integrates vulnerability manifestation localization, path-sensitive exploration, open-weight model specialization, and symbolic reasoning into a unified pipeline (\S\ref{sec:approach}).

\item 
We evaluate {\tech} on 490 benchmark vulnerabilities and 250 real-world CVEs without public PoVs, showing strong PoV-generation success 
and cost-effectiveness (\S\ref{sec:rq1}--\S\ref{sec:rq5}).

\item We demonstrate practical impact by identifying six flawed patches and five previously unreported vulnerabilities, four of which have been confirmed and fixed (\S\ref{sec:useful}).
\end{itemize}

\begin{figure}[tp]
 \vspace{0pt}
 \centering
    \includegraphics[width=0.7\linewidth]{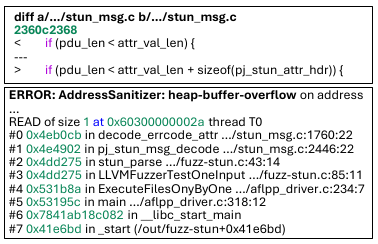}
    \vspace{-10pt}
    \caption{An example of patch (code diff at top) and sanitizer output (log at bottom) for the CVE-2022-23537 vulnerability from the PJSIP project.}
    \label{fig:motivation}
    \vspace{-15pt}
\end{figure}

\section{Background and Motivation}\label{sec:bkgmot}


\vspace{-2pt}
\subsection{CVE–PoV Gap: A Motivating Study}
\label{sec:cves}
\vspace{-2pt}

The national vulnerability database (NVD)~\cite{nvd} records publicly disclosed vulnerabilities. While patches are frequently available, concrete PoVs—inputs that reliably trigger the vulnerable behavior—are often missing. 


To quantify this gap, we analyze \emph{7,474} CVEs related to open-source projects from 2012 to 2024. 
We found that 
73.0\% of these vulnerabilities have patches available, yet only 16.8\% have public PoVs. The ExploitDB dataset~\cite{exploitdb} exhibits an even larger gap: among 5,412 CVEs in the same time period, only 131 (2.42\%) of them have PoVs available. 

This disparity highlights a fundamental limitation in current vulnerability disclosure practices: while patches indicate that a vulnerability exists, the absence of PoVs prevents reproducible validation of the vulnerability and its fix. Bridging this gap requires automated techniques that is capable of generating PoVs 
from disclosed vulnerabilities.

\vspace{-2pt}
\subsection{Challenges and Design Insights}
\label{sec:challenges}
\vspace{-2pt}

Generating PoVs for disclosed vulnerabilities differs from general testing: the goal is not broad coverage, but synthesizing a concrete input that triggers a specific vulnerable behavior, 
often without any relevant existing test (seed) inputs. 
This setting raises three challenges that motivate {\tech}'s design.

\begin{itemize}[leftmargin=*,itemsep=0pt, topsep=0pt]

\item \textbf{Challenge 1: vulnerability-targeted exploration.}
Only a small fraction of program paths can trigger a given vulnerability, so unguided fuzzing or symbolic exploration can spend most effort on irrelevant paths. Even when a patch is available, the patched statement may not coincide with where the vulnerability manifests~\cite{li2024effectiveness}. For example, in Figure~\ref{fig:motivation}, the patch is located (at line~2360) far from where the vulnerability is manifested (at line~1760), making the patch location alone an imprecise exploration target.

\textbf{\ul{Insight 1: semantic focusing.}}
{\tech} narrows the search \textit{before} constraint reasoning by localizing the vulnerability manifestation point. Anchoring exploration at this point turns broad program exploration into a focused reachability problem from program entries. 

\begin{figure*}[tp]
\centering
\vspace{0pt}
    \includegraphics[width=0.97\linewidth]{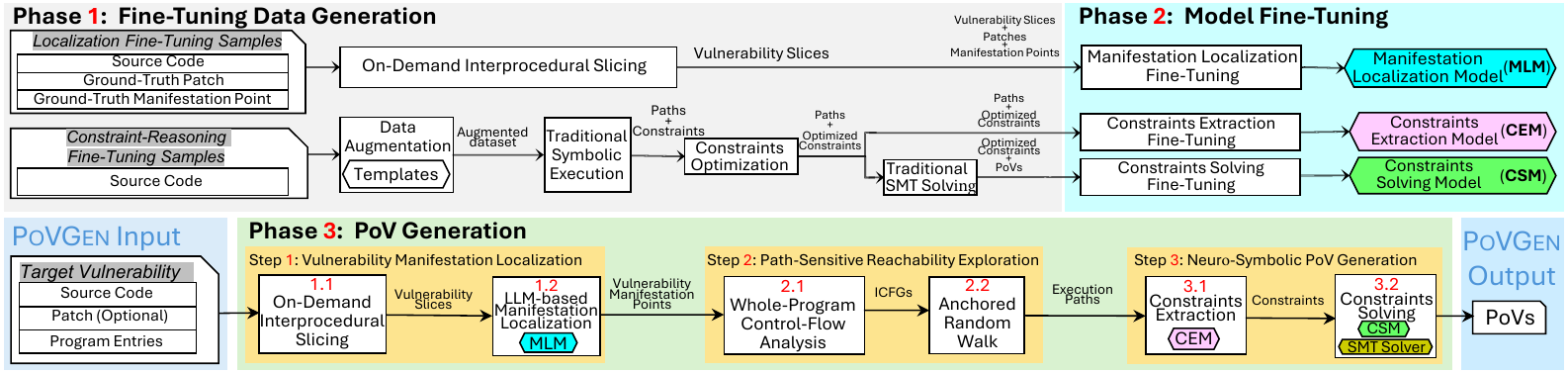}
        \vspace{-8pt}
        \caption{An overview of {\tech}'s design, including its input, three main working phases (and steps), and output.}
        \label{fig:overview}
        \vspace{-12pt}
\end{figure*}

\item \textbf{Challenge 2: constraint recovery for real programs.}
After relevant paths are identified, generating a PoV still requires constraints that capture how inputs drive execution to the vulnerable state. These constraints are difficult to recover in practice because they often span multiple functions and may depend on program state, pointer relationships, input-format rules, and external library behavior and environment.

\textbf{\ul{Insight 2: symbolic reasoning with model-guided constraint synthesis.}}
{\tech} combines symbolic program analysis with fine-tuned open-weight models. One model extracts constraints from the relevant paths, and another synthesizes inputs that satisfy them. An SMT solver is used as a fallback 
for constraints that 
can be solved directly but the model fails to solve. 
This keeps model reasoning focused on structured sub-tasks rather than asking a general-purpose model to infer an entire PoV from raw program context.

\item \textbf{Challenge 3: cost and controllability.}
PoV generation requires repeated reasoning over many candidate paths. Directly invoking frontier proprietary LLMs for every path can be expensive, and model behavior, availability, and pricing remain outside the user's control (e.g., commercial models 
can be retired or restricted at vendor's discretion). For example, reproducing one CVE in the PJSIP project (Figure~\ref{fig:motivation}) involves analyzing 106.8k code lines and consumes an average of 4.57M tokens over all explored paths. 

\textbf{\ul{Insight 3: open-weight task specialization.}}
{\tech} specializes open-weight models for vulnerability localization, constraint extraction, and constraint-guided input generation. The model artifacts can be pinned, audited, and reused across experiments, while avoiding per-sample proprietary API calls in the core pipeline. This trades general-purpose flexibility for lower cost, greater controllability, task-specific output regularity, and long-term stability and availability.

\end{itemize}

\vspace{-0pt}
\section{Approach}\label{sec:approach}
\vspace{-2pt}
We present our technique, starting with the design overview followed by details on each component. 

\vspace{-2pt}
\subsection{Overview}
\vspace{-2pt}


Figure~\ref{fig:overview} shows our {\tech} design. We target the realistic setting of automatically reproducing a disclosed vulnerability where the source code, program entries, and optionally the patch are available. As discussed in \S\ref{sec:bkgmot}, PoV generation 
needs to 
identify vulnerability-relevant paths and construct triggering inputs (without 
existing
seeds). {\tech} addresses this 
with a two-level design: an \textit{offline specialization} stage that prepares task-specific models, and an \textit{online PoV-generation} stage that uses these models to perform semantic focusing, path-sensitive exploration, and neuro-symbolic input generation.

\textbf{Phase 1 (Fine-Tuning Data Generation).} The \textit{offline specialization} stage has two phases. Phase~1 prepares task-specific datasets via two pipelines: on-demand interprocedural slicing for localization, and symbolic-execution/SMT-derived paths (augmented with templates) for constraint reasoning. \textbf{Phase 2 (Model Fine-Tuning).} It then trains three specialized models---the Manifestation Localization Model (MLM), Constraints Extraction Model (CEM), and Constraints Solving Model (CSM)---decomposing the task so each model specializes without interference. \textbf{Phase 3 (PoV Generation).} The \textit{online} stage applies these models in three steps: Step~1 localizes the manifestation point with the MLM over on-demand slices; Step~2 runs an Anchored Random Walk on the interprocedural control-flow graph to extract viable candidate paths; and Step~3 uses the CEM to derive SMT constraints per path and the CSM (with an SMT-solver fallback) to solve them into a PoV, augmenting solver-based reasoning with semantic inference for paths that would stall traditional symbolic execution. We detail each phase next.

To accommodate different scenarios, {\tech} operates in \ul{two modes}. In the \texttt{patch-guided mode}, the patch serves as a semantic anchor, enabling precise manifestation localization and a focused, multi-segment search (Entry $\rightarrow$ Patch $\rightarrow$ Manifestation). In the \texttt{patch-free} mode, {\tech} relies solely on the source code, predicting the manifestation point directly and exploring paths from the entry point. This mode acts as a fallback, trading off some analysis effectiveness for the ability to handle unpatched vulnerabilities.

\vspace{-2pt}
\subsection{Fine-Tuning Data Generation (Phase 1)}
\vspace{-2pt}

Fine-tuning is adopted as PoV generation requires structured outputs (manifestation locations, SMT constraints, and concrete satisfying inputs) where general-purpose (open-weight) LLMs hallucinate or violate formal syntax often. {\tech} uses two pipelines: one for MLM, and one for CEM and CSM.

\subsubsection{Localization Fine-Tuning Data Generation} \label{sec:slicing2}

{\tech} fine-tunes the MLM to identify vulnerability manifestation points, which serve as exploration targets. As summarized in Algorithm~\ref{algo:patchguidedscoping}, it performs on-demand interprocedural slicing to produce focused code slices that---paired with ground-truth patches and manifestation points (\(MPs\))---fine-tune the MLM.

The goal of slicing here is to construct a compact yet semantically sufficient context for manifestation localization. For training samples with ground-truth patches, {\tech} starts from the patch location, based on the insight that \textit{the manifestation point is often data- or control-dependent on the patched code}.
For each sample program, {\tech} first builds a system dependence graph (SDG) through interprocedural dependency analysis. The SDG combines the program dependence graphs (PDGs) of individual functions and adds interprocedural control- and data-dependence edges~\cite{horwitz1990interprocedural}, providing the basis for slicing.

A whole-program slice is too large for the MLM input context. {\tech} thus performs \textbf{On-Demand Interprocedural Slicing}: it starts from the patching function (or program entries when patch is unavailable), applies forward and backward slicing to retain statements dependent on the patching statements, and incrementally expands the slice with additional context functions. To identify which context functions are needed, {\tech} prompts a general-purpose, open-weight LLM with the current slice and asks it to identify additional functions whose semantics are necessary for localization~\cite{shahandashti2024program}. The SDG is then used to expand the slice accordingly.

This expansion proceeds iteratively until a threshold is reached (Lines~\ref{a1-line13}--\ref{a1-line15}). The threshold balances context completeness and tractability: it allows {\tech} to recover interprocedural dependencies without exceeding LLM context limits. We set it to three based on statistics from InterPVD~\cite{li2024effectiveness} and ARVO~\cite{mei2024arvo}, where most manifestation-relevant functions are reachable within three expansion rounds. The resulting triples of code slices, ground-truth patches, and ground-truth manifestation points form the fine-tuning data for the MLM.

\setlength{\textfloatsep}{0.05cm}
\begin{algorithm}[tp]
\scriptsize
\caption{Manifestation Localization Fine-Tuning}
\label{algo:patchguidedscoping}
\SetKwProg{Fn}{Function}{}{end}
\SetKwFunction{OnDemandSlicing}{OnDemandSlicing}
\SetKwFunction{MLMFineTuning}{MLMFineTuning}
\SetKwFunction{ConstructSDG}{ConstructSDG}
\SetKwFunction{GetPatchLocation}{GetPatchLocation}
\SetKwFunction{InitialSlice}{InitialSlice}
\SetKwFunction{QueryLLMForContext}{QueryLLMForContext}
\SetKwFunction{ExpandSlice}{ExpandSlice}
\SetKwFunction{Predict}{Predict}
\SetKwFunction{FineTune}{FineTune}
\SetKw{Return}{return}
\LinesNumbered
\KwIn{\( D_{ft} \): Localization fine-tuning samples}
\KwOut{\( MLM \): Manifestation Localization Model (MLM)}
\Fn{\MLMFineTuning{\(D_{ft}\)}}{
    $Progs, Patches, MPs \gets D_{ft}$ \\
    $Slices \gets$ []; \hfill  { \label{a1-line2}} \\
    \ForEach{$(P_i, Pa_i) \in (Progs, Patches)$\label{a1-line3}}{
        \(\mathcal{S}_i\) $\gets$ \OnDemandSlicing{\(P_i, Pa_i\)}; \hfill \\
        $Slices$.insert($\mathcal{S}_i$); \label{a1-line5}
    }
    $MLM \gets$ \FineTune{\(Slices,Patches,MPs\)}; \hfill 
    \\ \label{a1-line6}
    \Return \(MLM\)\;\label{a1-line9}
}
\Fn{\OnDemandSlicing{\(P, Pa\)}}{\label{a1-line10}
    $SDG$ $\gets$ \ConstructSDG{\(P\)}; \\ 
    $l_{patch}$ $\gets$ \GetPatchLocation{\(Pa\)}; \\
    \(\mathcal{S}\) $\gets$ \InitialSlice{$SDG, l_{patch}$}; \\
    \For{$i \gets 1$ \KwTo $\text{MAX\_ITER\_MLM}$\label{a1-line13}}{
        $F_{context}$ $\gets$ \QueryLLMForContext{\(\mathcal{S}\)}; \\
        \(\mathcal{S} \gets \mathcal{S} \cup \) \ExpandSlice{\(SDG, \mathcal{S}, F_{context}\)}; \label{a1-line15}
    }
    \Return \(\mathcal{S}\)\; \label{a1-line16}
}
\end{algorithm}

\begin{figure*}[tp]
\centering
\vspace{0pt}
	\includegraphics[width=0.9\linewidth]{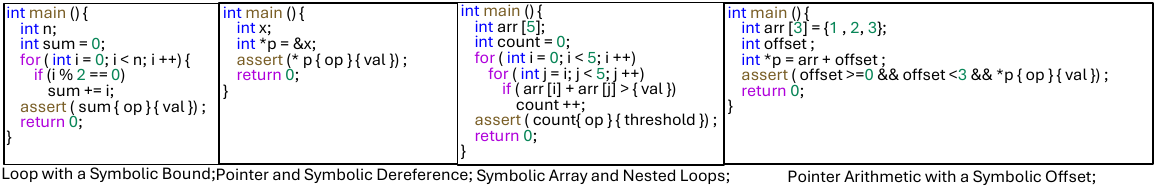}
        \vspace{-10pt}
        \caption{The templates used for data augmentation, where \{op\} is a random relational operator, \{val\} and \{threshold\} are random integers.}
	\label{fig:template}
        \vspace{-10pt}
\end{figure*}

\subsubsection{Constraint-Reasoning Fine-Tuning Data Generation}

To support LLM-guided constraint reasoning, we prepare training data for two separate models: 
CEM, which extracts constraints from given paths, and CSM, which generates satisfying inputs from constraints. Algorithm~\ref{algo:neuralsymbolic_detailed} summarizes this process. It augments the base training set with predefined templates (Lines~\ref{a3-line3}--\ref{a3-line4}), applies symbolic execution and SMT solving to generate path constraints and PoVs (Lines~\ref{a3-line5}--\ref{a3-line11x}), and uses the resulting pairs to fine-tune CEM and CSM 
(Lines~\ref{a3-line12}--\ref{a3-line13}).



{\tech} starts from fine-tuning samples collected from existing static-analysis benchmarks, which symbolic execution engines can process to generate paths, constraints, and satisfying inputs. These benchmarks provide diverse vulnerability patterns and much of the real-world complexity needed for training. However, they underrepresent several control- and data-flow motifs common in real CVEs, such as loops with symbolic bounds and pointer arithmetic with symbolic offsets. Without such motifs, the models are less prepared for the path-level reasoning required during PoV generation.

To complement the benchmark samples, {\tech} performs \textbf{Data Augmentation}, adding four templates (Figure~\ref{fig:template}) that inject these missing motifs: (1) \emph{loops with symbolic bounds}, (2) \emph{pointer and symbolic dereference}, (3) \emph{symbolic array and nested loops}, and (4) \emph{pointer arithmetic with a symbolic offset}. These templates are not intended to cover the full diversity of real-world CVEs; that diversity comes primarily from the benchmark samples. Instead, they expose the models to patterns that frequently cause path explosion, memory aliasing, and complex constraints in symbolic execution. Together, the benchmark samples and templates improve the models' ability to handle real-world control- and data-flow structures.


{\tech} then runs a \textbf{Traditional Symbolic Execution} engine on the augmented sample programs. Variables that influence vulnerability triggering and appear at program entry points are marked symbolic, such as loop bounds, arrays, and pointer offsets in the templates. For each explored path from a program entry to an exit point, the engine accumulates logical constraints over the symbolic variables.



While the raw constraints from the symbolic execution engine are logically valid, they are often unsuitable for fine-tuning because they contain redundant or noisy expressions (e.g., trivially true clauses like ``x == x"). 
Thus, {\tech} performs \textbf{Constraint Optimization} to \textit{normalize and simplify} constraints using a few common rules:

\begin{itemize}[leftmargin=*,itemsep=0pt, topsep=0pt]
    \item \textit{Constant Folding.} Replace constant-only sub-expressions with their values, e.g., \texttt{(+ 2 2)} 
    is simplified as 
    \texttt{4}.
    \item \textit{Algebraic Simplification.} Apply algebraic identities, e.g., \texttt{(* x 1)} becomes (is simplified as) \texttt{x}.
    \item \textit{Logical Simplification.} Simplify Boolean expressions, e.g., \texttt{(and p true)} becomes (is simplified as) \texttt{p}.
    \item \textit{Canonical Reordering.} Reorder commutative expressions into a canonical form, e.g., \texttt{(+ y x)} becomes 
    \texttt{(+ x y)}.
\end{itemize}



For each valid set of optimized path constraints, {\tech} performs \textbf{Traditional SMT Solving}: it 
queries an SMT solver to obtain concrete input values satisfying the constraints. The satisfying assignment is treated as a training PoV for that path. 

Finally, 
(\textit{paths, optimized constraints}) and (\textit{optimized constraints, PoVs}) pairs form the fine-tuning data for the two constraint-reasoning models (CEM and CSM), respectively.

\vspace{-2pt}
\subsection{Model Fine-Tuning (Phase 2)}
\vspace{-2pt}


Using the datasets from Phase 1, {\tech} fine-tunes three specialized models for manifestation localization and constraint reasoning. As motivated in \S\ref{sec:bkgmot}, fine-tuning transfers the structured reasoning patterns of symbolic execution and SMT solving into the models---closing the distributional gap with their pretraining---while local deployment (of the fine-tuned, \textit{open-weight} models) keeps marginal inference cost low.

\subsubsection{Manifestation Localization Fine-Tuning}\label{sec:localization-fine-tuning}



The MLM is fine-tuned to identify the vulnerability manifestation statement within a given code slice, enabling semantic focusing in subsequent analysis. We formulate this as an instruction-following task, where the input is a code slice and the output is the line corresponding to the manifestation point.

The training data consists of slices generated in Phase 1, paired with ground-truth manifestation points. To improve robustness, we include negative samples: if a slice does not contain the manifestation point, the expected output is \texttt{None}. This teaches the model to reject irrelevant contexts and reduces false positives. When none of the generated slices contains the manifestation point, we construct a ground-truth slice spanning from the patch location to the manifestation point to ensure at least one positive example.

%

We fine-tune the model by optimizing the parameters $\theta$ to minimize the negative log-likelihood of the correct output $Y$ given the input instruction $I$, i.e., $\mathcal{L} = -\log P(Y \mid I, \theta)$.

\subsubsection{Constraint-Reasoning Fine-Tuning}

\begin{algorithm}[t]
\scriptsize
\caption{Constraint-Reasoning Fine-Tuning}
\label{algo:neuralsymbolic_detailed}
\SetKwProg{Fn}{Function}{}{end}
\SetKwFunction{NeuralSymbolicFineTuning}{NeuroSymbolicFineTuning}
\SetKwFunction{GenerateFromTemplates}{GenerateFromTemplates}
\SetKwFunction{SymbolicExecute}{SymbolicExecute}
\SetKwFunction{Optimize}{Optimize}
\SetKwFunction{FineTune}{FineTune}
\SetKwFunction{TradSMTSolver}{TradSMTSolver}
\SetKwFunction{Predict}{Predict}
\SetKwFunction{Solve}{Solve}
\SetKwFunction{BuildPoV}{BuildPoV}
\SetKw{Return}{return}
\LinesNumbered
\KwIn{\( D_{ft} \): neuro-symbolic Fine-Tuning Dataset}
\KwOut{\(CEM\): Constraints Extraction Model, \\ 
    \hspace{24pt} \(CSM\): Constraints Solving Model}
\Fn{\NeuralSymbolicFineTuning{\(D_{ft}\)}}{
    \( D_{aug} \gets \) \GenerateFromTemplates{}; \hfill 
    \\ \label{a3-line3}
    \( D_{neural} \gets D_{ft} \cup D_{aug}; \) \hfill 
    \\ \label{a3-line4}
    $\mathcal{P}aths \gets$ []; \\
    $Constraints \gets$ []; \hfill 
    \\ \label{a3-line5}
    $PoVs \gets$ []; \\
    \ForEach{\(P_i\) in \(D_{neural}\)}{
        \( constr \gets \) \SymbolicExecute{\(P_i\)}; \hfill \\
        \( constr_{opt} \gets \) \Optimize{\(constr\)}; \hfill \\
        \(pov \gets \) \TradSMTSolver{\(constr_{opt}\)} \\
        $\mathcal{P}aths$.insert($P_i$); \\
        $Constraints$.insert($constr_{opt}$); \label{a3-line9} \\
        $PoVs$.insert($pov$); \label{a3-line11x}\\
    }
    \(CEM \gets \) \FineTune{\(\mathcal{P}aths, Constraints\)}; \hfill 
    \\ \label{a3-line12}
    \(CSM \gets \) \FineTune{\(Constraints, PoVs\)};  \label{a3-line13}\hfill
    
    \Return \(CEM, CSM\)\; \label{a3-line18}
}
\end{algorithm}


{\tech} fine-tunes two models for neuro-symbolic PoV generation: the Constraints Extraction Model (CEM) and the Constraints Solving Model (CSM), following the process in Algorithm~\ref{algo:neuralsymbolic_detailed}.

The fine-tuning is structured as instruction-following tasks. The CEM learns to derive SMT constraints from given paths, mapping program semantics to logical formulas. The CSM learns to generate candidate inputs that satisfy these constraints, given constraint–PoV pairs produced in Phase 1.


Both models are trained by minimizing the negative log-likelihood of the target outputs given their respective inputs, following the same optimization formulation as in \S\ref{sec:localization-fine-tuning}. After fine-tuning, CEM and CSM are used in Phase 3 to support LLM-guided constraint reasoning during PoV generation.

\subsection{PoV Generation (Phase 3)}
\vspace{-2pt}

Given a target vulnerability, {\tech} generates a PoV in three steps that progressively narrow the search space and resolve constraint complexity, constructing triggering inputs from scratch without any seed.

\subsubsection{Vulnerability Manifestation Localization (Step 1)}

We apply the fine-tuned MLM to the target vulnerability to identify the manifestation point, which serves as the target for subsequent exploration. We first generate slices using the same \textbf{On-Demand Interprocedural Slicing} (1.1) process as in $\S$\ref{sec:slicing2}: in {\tt patch-guided} mode, slicing starts from the patched code lines; in {\tt patch-free} mode, it starts from program entries. 
This is followed by \textbf{LLM-based Manifestation Localization} (1.2): each resulting slice is fed into the MLM with the same instruction format as used during fine-tuning, and the model outputs the predicted manifestation point.

\subsubsection{Path-Sensitive Reachability Exploration (Step 2)}

This step identifies candidate paths from program entries to the manifestation point, optionally passing through the patch location (i.e., when available). As Algorithm~\ref{algo:pathsensitiveexploration} shows, {\tech} first constructs the interprocedural control-flow graph (ICFG) 
via whole-program control-flow analysis (Line~\ref{a2-line2}). Using the program entry ($E$), patch location ($L_{patch}$, if available), and manifestation point ($L_{manifest}$) as anchors, it then performs {anchored random walk}s on the ICFG to explore candidate control-flow paths $\mathcal{P}_{paths}$ as 
candidate paths 
(Lines~\ref{a2-line3}--\ref{a2-line7}). 

\vspace{-0pt}

\begin{algorithm}[tp]
\scriptsize
\caption{Path-Sensitive Reachability Exploration}
\label{algo:pathsensitiveexploration}
\SetKwProg{Fn}{Function}{}{end}
\SetKwFunction{PathSensitiveExploration}{PathSensitiveExploration}
\SetKwFunction{ConstructICFG}{ConstructICFG}
\SetKwFunction{RandomWalk}{RandomWalk}
\SetKw{Return}{return}
\LinesNumbered
\KwIn{\( P \): Target program, \( E \): Program entry, \( L_{patch} \): Patch location (optional), \( L_{manifest} \): Manifestation point}
\KwOut{\( \mathcal{P}_{paths} \): A set of viable candidate paths (exploration contexts)}
\Fn{\PathSensitiveExploration{\(P, E, L_{patch}, L_{manifest}\)}}{
    $ICFG \gets$ \ConstructICFG{\(P\)}; \hfill 
    \\ \label{a2-line2}
    $\mathcal{P}_{paths} \gets$ []; \\ \label{a2-line3}
    \For{$i \gets 1$ \KwTo \text{MAX\_ITER\_WALK} }{ \label{a2-line4}
        $path\gets$\RandomWalk{\(ICFG,E,L_{patch},L_{manifest}\)}; \hfill \\
        \If{$path$ is not null}{
            $\mathcal{P}_{paths}$.insert($path$); \hfill \\ 
            \label{a2-line7}
        }
    }
    \Return \(\mathcal{P}_{paths}\)\; \label{a2-line8}
}
\end{algorithm}


Specifically, the \textbf{Whole-Program Control-Flow Analysis} (2.1) 
first constructs the CFG for each function, followed by resolving calling relationships and return edges between functions.  
Finally, it builds the ICFG by connecting the function-level CFGs via the resolved call and return edges, providing a global view of control flow across the program.

\begin{algorithm}[tp]
\scriptsize
\caption{Neuro-Symbolic PoV Generation}
\label{algo:neuralsymbolic_detailed_old}
\SetKwProg{Fn}{Function}{}{end}
\SetKwFunction{NeuralSymbolicPoVGeneration}{NeuroSymbolicPoVGeneration}
\SetKwFunction{Predict}{Predict}
\SetKwFunction{Solve}{Solve}
\SetKwFunction{BuildPoV}{BuildPoV}
\SetKw{Return}{return}
\LinesNumbered
\KwIn{\( C_{explore} \): Exploration contexts from Step 2, \( desc \): Vulnerability description, \( patch \): Patch (optional)}
\KwOut{\( PoVs \): A set of generated Proof-of-Vulnerabilities}
\Fn{\NeuralSymbolicPoVGeneration{\(C_{explore}, desc, patch\)}}{
    $PoVs \gets$ []; \\ 
    \ForEach{\(path\) in \(C_{explore}\)}{ 
        \( constr \gets \) \Predict{\(CEM, path, desc, patch\)}; \hfill 
        \label{a4-line4} \\
        \( pov \gets \) \Predict{\(CSM, path, desc, patch, constr\)}; \hfill 
        \label{a4-line5} \\
        \If{pov is not valid}{ \label{a4-line6}
            \( pov \gets \) \Solve{\(SMTSolver, constr\)}; \hfill 
            \label{a4-line7} \\ 
        }
        $PoVs$.insert($pov$);
    }
    \Return \(PoVs\)\; \label{a3-line18}
}
\end{algorithm}

We then use \textbf{Anchored Random Walk} (2.2) to generate viable paths connecting the anchors.
%
Since enumerating all such paths is infeasible (path explosion) and purely random exploration is inefficient, the anchored walk biases exploration toward the anchors while preserving path diversity. 


Accordingly, we aim to discover paths that connect the program entry, the manifestation point, and, in {\tt patch-guided} mode, the patch location. The patch location is included because patched statements are directly related to the vulnerable program state; paths that traverse them are more likely to capture vulnerability-triggering conditions. These paths serve as the exploration context for the final PoV generation step.


\emph{Pre-computation phase.} 
We perform a backward breadth-first search (BFS) on the ICFG starting from each target anchor that the walk needs to reach. In {\tt patch-guided} mode, BFS is performed from both the patch location and the manifestation point; in {\tt patch-free} mode, the BFS is performed only from the manifestation point. This produces a distance map that records, for each reachable node of the ICFG, its shortest path distance to the target anchor. 
We use backward BFS because we need to know how far each node is from the target, not the other way around, and BFS is sufficient since ICFG edges are unweighted. These distance maps serve as the guidance signal during the subsequent exploration.

\emph{Exploration phase.} Starting from the program entry, the walk proceeds forward on the ICFG. At each branch, the next edge is selected randomly with probability inversely proportional to the successor's pre-computed distance to the current anchor (specifically, weight $w = 1/(d+1)$, where $d$ is the distance), so closer nodes are preferred but more distant nodes can still be chosen. This keeps the walk biased toward the anchor while retaining diversity. 

In {\tt patch-guided} mode, the walk is first guided by the distance to the patch location; once the patch location is reached, the guidance switches to the manifestation point. In {\tt patch-free} mode, the walk is guided directly toward the manifestation point. To ensure interprocedural search validity, we maintain a call stack: function-call edges push a return identifier onto the stack, and a return edge is only taken if its identifier matches the top of the stack. This prevents the walk from returning to an arbitrary caller (i.e., invalid returns). The walk terminates when the manifestation point is reached, when a maximum number of steps or restart attempts is exceeded, or when no valid successor exists.

To obtain diverse candidate paths, we run the anchored random walk multiple 
times with different random seeds (Lines~\ref{a2-line4}--\ref{a2-line7} in Algorithm~\ref{algo:pathsensitiveexploration}). The complete pseudocode is provided in the supplementary material. The resulting 
paths serve as the input for Step~3. We refer to these focused paths as the ``exploration context'', as they bridge the gap between high-level program structure (control flow) and the low-level details (constraint-level reasoning) required for PoV generation.


\subsubsection{Neuro-Symbolic PoV Generation (Step 3)}

The final step performs LLM-guided constraint reasoning to generate PoVs from the exploration contexts, as summarized in Algorithm~\ref{algo:neuralsymbolic_detailed_old}. 
For each execution path, {\tech} uses the fine-tuned CEM for \textbf{Constraint Extraction} (3.1), which derives SMT constraints that capture the path conditions, given the path, the vulnerability description, and the ground-truth patch when available (Line~\ref{a4-line4}). 
This is followed by \textbf{Constraint Solving} (3.2), where the fine-tuned CSM generates candidate inputs that satisfy these constraints in the same exploration context, hence producing a satisfying PoV (Line~\ref{a4-line5}).

To improve robustness, {\tech} incorporates an SMT solver as a fallback (Lines~\ref{a4-line6}--\ref{a4-line7}, also as seen in Figure~\ref{fig:overview}): when the CSM fails to produce a valid solution for complex or corner-case constraints, the SMT constraints are passed to the solver to obtain a satisfying solution. This combination of LLM-guided reasoning and SMT fallback enables {\tech} to handle constraints that are difficult for either approach alone.

Notably, 
we instruct CSM to synthesize a program (Python script) 
that generates the PoV rather than the PoV itself directly. 
This avoids ambiguity in representing raw bytes, escape sequences, and structured binary fields in natural-language output, and makes each generated PoV directly reproducible as an executable artifact.
This step addresses the complexity of vulnerability-triggering conditions by combining semantic inference with formal constraint solving, enabling PoV generation for paths that would cause traditional symbolic execution to time out or fail due to incomplete constraint modeling.

\section{Implementation}






For vulnerability manifestation localization, we use Joern~\cite{joern} to construct PDGs and call graphs, and adapt SySeVR~\cite{li2021sysevr} for slicing. Context functions for on-demand interprocedural slicing are identified using GLM-5.2~\cite{glm52}, an open-weight model, via its API. 
We set the slicing depth to three, following prior empirical findings~\cite{li2024effectiveness}.

For path-sensitive reachability exploration, we compile target programs with WLLVM~\cite{wllvm} and construct 
ICFGs 
using SVF~\cite{sui2016svf}. We implement the anchored random walk via SVF API and extract up to 200 candidate paths in parallel.

The rationale of these choices is 
Joern offers efficient source-level 
slicing, while SVF offers precise interprocedural control-flow modeling. For scalability on large codebases we use only SVF's lightweight control-flow analysis, balancing efficiency against the path accuracy PoV generation requires.

For neuro-symbolic constraint reasoning, we generate path constraints and corresponding satisfying inputs using symbolic execution engine KLEE~\cite{cadar2008klee} and SMT solver Z3~\cite{de2008z3}. These artifacts are used to fine-tune the CEM and CSM. Model fine-tuning is performed with Unsloth~\cite{han2024unsloth}, using Llama-3.2-3B for MLM and Llama-3.1-8B for both CEM and CSM.

\textit{No component of {\tech} depends on closed, commercial LLMs, so it can be reproduced with publicly available models (and self-hosted where API access is undesirable)}.




\section{Evaluation}



\noindent
We evaluate {\tech} around five research questions (RQs): 

\begin{itemize}[leftmargin=*, noitemsep,topsep=1pt, itemindent=-24pt]
\item
\textbf{RQ1.} How effective is {\tech}?
\item
\textbf{RQ2.} How does {\tech} compare to 
different
baselines?
\item
\textbf{RQ3.} How does each component of {\tech} contribute? 
\item
\textbf{RQ4.} 
What runtime and model cost does {\tech} incur?
\item
\textbf{RQ5.} Can {\tech} reproduce real-world CVEs?
\end{itemize}




\subsection{Datasets and Setup} 
\vspace{-3pt}
We use three datasets for fine-tuning and evaluation.

\textbf{Localization Fine-Tuning Dataset.} 
For MLM fine-tuning, we use the InterPVD dataset~\cite{li2024effectiveness}, which provides labeled patches and manifestation points for 769 C/C++ vulnerabilities across 53 projects and 16 CWEs.

\textbf{Constraint-Reasoning Fine-Tuning Dataset.} 
For CEM and CSM, we collect 484 samples from SV-COMP~\cite{beyer2021software} and 15,500 (memory-safety) samples from the Juliet Test Suite~\cite{black2018juliet}. We further augment these with 10,000 synthetic samples generated from code templates (Figure~\ref{fig:template}), resulting in 25,984 samples for constraint-reasoning fine-tuning.

\textbf{Evaluation Datasets.} 
%
%
%
For RQ1--RQ4, we use ARVO~\cite{mei2024arvo}, which contains 5,000 memory vulnerabilities with ground-truth PoVs across 250+ C/C++ projects. We randomly sample 490 cases, for a 98\% confidence level with $\pm$5\% margin of error over the full population~\cite{cochran1977sampling}. These 490 cases span 110 projects and 10 vulnerability categories, as shown in Table~\ref{tab:effectiveness1}.
For RQ5, we collect vulnerabilities from the CVE database (2012--2024) with available patches but no public PoVs. To ensure a controlled evaluation, we restrict to projects included in ARVO, which provides reproducible build environments and harnesses. This prevents environment setup failures from being conflated with PoV-generation failures. 
Both evaluation datasets are disjoint from the fine-tuning data. The constraint-reasoning fine-tuning data (SV-COMP and Juliet) contains no CVEs, so it is disjoint from ARVO and the CVE set by construction. For localization, we cross-check both ARVO and the collected CVEs against the InterPVD dataset~\cite{li2024effectiveness}: ARVO contains no overlapping case, and we exclude the 19 CVEs that also appear in InterPVD. The resulting 250 CVEs span 62 projects and 14 CWEs, as shown in Table~\ref{tab:cve-results}, and no evaluation sample appears in any fine-tuning dataset.



\textbf{Scope.}
We focus on PoV generation (i.e., generating inputs that trigger vulnerabilities). Automated environment setup and harness construction are outside our scope; thus, we use ARVO’s pre-built environments to enable scalable evaluation. Our evaluation targets C/C++ memory-safety vulnerabilities due to their prevalence and impact, as well as the availability of high-quality benchmarks. 

\textbf{Environment.}
All experiments run on a workstation with an AMD Threadripper Pro 5595WX CPU, four NVIDIA RTX A6000 GPUs, and 512GB RAM, running Ubuntu 22.04.

\begin{table}[t]
 \centering
  \vspace{-0pt}
 \caption{Effectiveness of \textsc{PoVGen} (with/without patch information) vs.\ baselines on the ARVO dataset (\#samples in parentheses)}
 \vspace{-5pt}
 \label{tab:effectiveness1}
\scalebox{0.72}{
\begin{tabular}{l@{\hspace{6pt}}r@{\hspace{6pt}}r@{\hspace{8pt}}r@{\hspace{6pt}}r@{\hspace{6pt}}r@{\hspace{6pt}}r}
\hline
 & \multicolumn{2}{c}{\textbf{\textsc{PoVGen}}} & \multicolumn{4}{c}{\textbf{Baselines}} \\
\cline{2-3}\cline{4-7}
\textbf{Vulnerability Type} & \textbf{w/ Patch} & \textbf{w/o Patch} & \textbf{LibFuzzer} & \textbf{AFL++} & \textbf{AFLGo} & \textbf{KLEE} \\
\hline
SEGV (55) & \textbf{72.73\%} & 65.45\% & 50.91\% & 47.27\% & 36.36\% & 3.64\% \\
global-buffer-overflow (22) & \cellcolor{gray!15}\textbf{90.91\%} & 63.64\% & 50.00\% & 54.55\% & 45.45\% & 0.00\% \\
heap-buffer-overflow (243) & \textbf{80.66\%} & 65.02\% & 49.79\% & 36.63\% & 45.27\% & 2.06\% \\
heap-use-after-free (35) & \textbf{77.14\%} & 62.86\% & 57.14\% & 42.86\% & 54.29\% & 2.86\% \\
stack-buffer-overflow (31) & \textbf{74.19\%} & 67.74\% & 35.48\% & 54.84\% & 48.39\% & 0.00\% \\
stack-buffer-underflow (3) & \cellcolor{gray!15}\textbf{100.00\%} & \underline{33.33\%} & 33.33\% & 0.00\% & 33.33\% & 0.00\% \\
stack-use-after-return (3) & 66.67\% & 66.67\% & 66.67\% & 33.33\% & \textbf{100.00\%} & 0.00\% \\
undefined-behavior (16) & \textbf{56.25\%} & 37.50\% & 37.50\% & 50.00\% & 31.25\% & 6.25\% \\
use-after-poison (7) & \textbf{85.71\%} & 71.43\% & 28.57\% & 71.43\% & 71.43\% & 0.00\% \\
use-of-uninitialized-value (75) & \textbf{81.33\%} & \cellcolor{gray!20}\textbf{72.00\%} & 58.67\% & 34.67\% & 40.00\% & 4.00\% \\
\hline
\rowcolor{gray!25}
\textit{Overall (490)} & \textit{\textbf{78.98\%}} & \textit{65.10\%} & \textit{50.20\%} & \textit{40.61\%} & \textit{44.49\%} & \textit{2.45\%} \\
\hline
\end{tabular}
}
 \vspace{4pt}
\end{table}

\subsection{RQ1: Effectiveness}\label{sec:rq1}

To evaluate whether {\tech} can reproduce vulnerabilities, we generate a concrete PoV either by executing the Python script produced by the CSM or by using the satisfying assignment returned by the SMT solver fallback. We then run the target harness with the generated PoV as input. A reproduction is considered \textit{successful} only if the sanitizer reports an error whose vulnerability type and crash location match those triggered by the ground-truth PoV. This strict criterion ensures that {\tech} reproduces the intended vulnerability rather than merely crashing the program.

\begin{figure}[tp]
\centering
\vspace{-0pt}
	\includegraphics[width=1.0\linewidth]{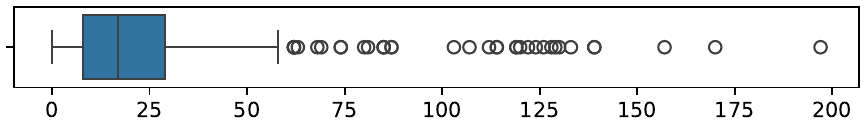}
        \vspace{-22pt}
        \caption{Distribution of \#execution paths explored to generate success PoVs.}
	\label{fig:path-boxplot}
        \vspace{-10pt}
\end{figure}
\begin{figure}[tp]
\centering
\vspace{-2pt}
	\includegraphics[width=1.0\linewidth]{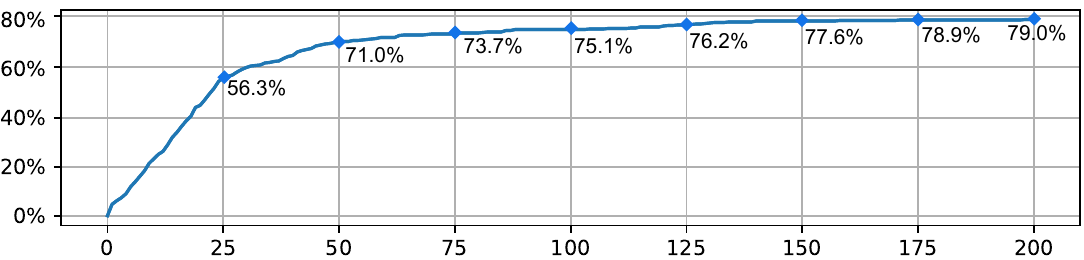}
        \vspace{-22pt}
        \caption{Success rate of {\tech} in terms of number of paths explored.}
	\label{fig:path-succ}
        \vspace{4pt}
\end{figure}

Table~\ref{tab:effectiveness1} reports {\tech}'s effectiveness on ARVO samples in \texttt{patch-guided} mode. Overall, {\tech} reproduces 78.98\% of the vulnerabilities, with success remaining high across most categories, showing that manifestation localization, path-sensitive exploration, and neuro-symbolic PoV generation jointly reproduce diverse memory-safety vulnerabilities. Examining how many paths are needed (Figure~\ref{fig:path-boxplot}), the distribution is heavily skewed toward small values---median approximately 20 paths, with 75\% of successful reproductions within the first 30---and the cumulative success rate (Figure~\ref{fig:path-succ}) reaches 71.0\% within 50 explored paths and then plateaus, indicating that {\tech}'s anchored exploration quickly concentrates on vulnerability-relevant paths.


To evaluate the \texttt{patch-free} mode, we remove patch information from ARVO samples to simulate settings where no patch is available. As Table~\ref{tab:effectiveness1} shows, {\tech} still reproduces 65.10\% of the vulnerabilities. The reduction from \texttt{patch-guided} mode is mainly due to Step~1: without patch information, manifestation localization recall decreases from 79.18\% to 62.65\% (as shown in Table~\ref{tab:loc-acc}). Nevertheless, the remaining stages continue to generate PoVs from less precise anchors, showing that {\tech} remains effective even when semantic guidance is weaker.

\begin{table}[tp]
 \centering
 \vspace{-8pt}
 \caption{Manifestation localization effectiveness}
 \vspace{-6pt}
 \label{tab:loc-acc}
\scalebox{0.9}{
\begin{tabular}{l@{\hspace{10pt}}r@{\hspace{10pt}}r@{\hspace{10pt}}r}
\hline
\textbf{Approach} & \textbf{Recall} & \textbf{Precision} & \textbf{F1} \\
\hline
\textsc{PoVGen} & \cellcolor{gray!20}\textbf{79.18\%} & \cellcolor{gray!20}\textbf{58.23\%} & \cellcolor{gray!20}\textbf{67.11\%} \\
\textsc{PoVGen-NoPatch} & 62.65\% & 38.98\% & 48.06\% \\
VulTrigger & \underline{30.63\%} & \underline{14.67\%} & \underline{19.83\%} \\
Gemini-2.5-Flash & 55.25\% & 43.79\% & 48.86\% \\
Llama-3.2 & 32.68\% & 22.59\% & 26.71\% \\
\hline
\end{tabular}
}
\vspace{6pt}
\end{table}

\subsection{RQ2: Comparison to Existing Techniques}\label{sec:rq2}
\textbf{Comparison to Fuzzers.} 
%
We compare {\tech} against SOTA fuzzers---LibFuzzer~\cite{serebryany2016continuous}, AFL++~\cite{fioraldi2020afl++}, and the directed fuzzer AFLGo~\cite{bohme2017directed}---using each ARVO project's official seed corpus and configuring AFLGo with patch locations as targets; each experiment is repeated three times (union of reproduced vulnerabilities) with a one-week timeout. As shown in Table~\ref{tab:effectiveness1}, the fuzzing baselines reach at most 50.20\% success: despite ARVO vulnerabilities originally being found via fuzzing (OSS-Fuzz), reproduction remains hard without targeted path exploration, whereas {\tech}'s semantic focusing and guided exploration reach triggering paths more effectively.

\textbf{Comparison to Symbolic Execution Engine.} 
%
%
We then compare {\tech} with KLEE~\cite{cadar2008klee}, using each sample's LLVM bitcode, marking inputs symbolic, stubbing sanitizer functions, and enabling all of KLEE's memory-safety checkers (three runs, one-week timeout). As shown in Table~\ref{tab:effectiveness1}, KLEE reproduces only 2.45\%, reflecting well-known limitations on real-world code: path explosion restricts coverage; external/system calls and incomplete models hinder exploration; and the flat byte-level memory model struggles with symbolic pointer arithmetic and aliasing. In contrast, {\tech} avoids exhaustive exploration by focusing on vulnerability-relevant paths and combining semantic inference with constraint reasoning.
Notably, although fuzzing and KLEE use no patch information, {\tech} outperforms them even in its {\tt patch-free} mode (Table~\ref{tab:effectiveness1}).

\begin{table}[tp]
 \centering
 \vspace{0pt}
 \caption{Effectiveness of \textsc{PoVGen} compared to 
 direct frontier-LLM prompting 
 on the partial (context-fitting) ARVO dataset}
 \vspace{-8pt}
 \label{tab:effectiveness2}
\scalebox{0.9}{
\begin{tabular}{l@{\hspace{8pt}}r@{\hspace{8pt}}r@{\hspace{8pt}}r@{\hspace{8pt}}r}
\hline
\textbf{Vulnerability Type} & \textbf{PoVGen} & \textbf{Gemini-2.5} & \textbf{OpenAI-o3} & \textbf{Claude-4} \\
\hline
SEGV  & \textbf{83.33\%} & 50.00\% & 33.33\% & 16.67\% \\
global-buffer-overflow & \textbf{75.00\%} & 25.00\% & 25.00\% & 25.00\% \\
heap-buffer-overflow & \textbf{78.57\%} & 21.43\% & 35.71\% & 17.86\% \\
heap-use-after-free & \textbf{66.67\%} & \textbf{66.67\%} & \textbf{66.67\%} & \textbf{66.67\%} \\
use-of-uninitialized-value & \textbf{77.78\%} & 55.56\% & 33.33\% & 22.22\% \\
\hline
\rowcolor{gray!25}
\textit{Overall} & \cellcolor{gray!35}\textit{\textbf{78.00\%}} & \textit{34.83\%} & \textit{35.89\%} & \textit{22.01\%} \\
\hline
\end{tabular}
}
\vspace{-6pt}
\end{table}

\textbf{Comparison to Direct Frontier-LLM Prompting.}
%
%
%
We also compare {\tech} with 
frontier LLMs: OpenAI-o3, Claude-Sonnet-4, and Gemini-2.5-Pro. 
To enable 
diagnostic 
comparisons, 
for each baseline LLM we (1) provide the CVE description, patch, and program source code of each sample 
and (2) use a structured multi-step prompt that mirrors {\tech}'s workflow. 
Specifically, the prompt instructs the LLM to be an expert cybersecurity researcher and to solve the problem by first identifying the vulnerability manifestation point, then formulating a viable candidate path to that point, next deriving the necessary input constraints for that path, and finally generating a PoV script. The 
prompt template detailing this structured task is provided in the supplementary material. 


\begin{table}[tp]
 \centering
 \vspace{-2pt}
  \caption{Direct accuracy evaluation of CEM and CSM on the partial (context-fitting) ARVO dataset}
 \vspace{-8pt}
 \label{tab:direct-accuracy}
\scalebox{0.7}{
\begin{tabular}{lccccc}
\hline
 & \multicolumn{4}{c}{\textbf{Constraint Extraction (CEM)}} & \textbf{Solving (CSM)} \\
\cline{2-5}\cline{6-6}
 & \multicolumn{2}{c}{\textbf{Precision}} & \multicolumn{2}{c}{\textbf{Recall}} & \\
\cline{2-3}\cline{4-5}
\textbf{Approach} & \textbf{Syn.} & \textbf{Sem.} & \textbf{Syn.} & \textbf{Sem.} & \textbf{Success} \\
\hline
\rowcolor{gray!15}
\textbf{\textsc{PoVGen}} & \textbf{69.38\%} & \textbf{61.52\%} & \textbf{94.81\%} & \textbf{89.63\%} & \textbf{76.51\%} \\
Llama-3.1-8B & 20.11\% & 18.64\% & 60.74\% & 57.04\% & 37.96\% \\
Gemini-2.5-Flash & 60.42\% & 51.14\% & 87.32\% & 83.10\% & 67.96\% \\
Claude-Sonnet-4 & 68.72\% & 59.24\% & 92.52\% & 86.14\% & 73.46\% \\
Z3 only & -- & -- & -- & -- & 51.22\% \\
\hline
\end{tabular}
}
\vspace{4pt}
\end{table}


Due to the context window limits of these commercial LLMs (e.g., OpenAI-o3 and Claude-Sonnet-4 allow up to 200k tokens), we were only able to select 52 samples that fit. These samples include 6 SEGV, 4 global-buffer-overflow, 28 heap-buffer-overflow, 3 heap-use-after-free, and 11 use-of-uninitialized-value vulnerabilities. Considering the cost of these LLMs, we limited each to a maximum of 10 trials per sample if the generated PoVs failed to trigger the vulnerability. 

As shown in Table~\ref{tab:effectiveness2}, these LLM baselines achieve at most 35.89\% success, significantly below {\tech}'s 78.00\%. This gap highlights the limitation of general-purpose reasoning: without structured path exploration and carefully guided constraint reasoning, LLMs struggle to produce valid PoVs even when instructed via detailed, reasoning-oriented prompts.


\begin{table*}[tp]
 \centering
     \caption{How each key component of {\tech} contributes to its performance} 
     \vspace{-8pt}
\scalebox{0.62}{
\begin{tabular}{lc|ccccccccc}
\hline
      &       & \multicolumn{9}{c}{\textbf{Ablated Version of PoVGen}} \\ \hline
\textbf{Vulnerability Type} & \multicolumn{1}{l|}{\textbf{PoVGen}} & \multicolumn{1}{l}{\textbf{No SMT Solver}} & \multicolumn{1}{l}{\textbf{No Simp}} & \multicolumn{1}{l}{\textbf{No Simp\&Aug}} & \multicolumn{1}{l}{\textbf{Non-Fine-Tune/Powerful S3}} & \multicolumn{1}{l}{\textbf{No-Fine-Tune/Base S3}} & \multicolumn{1}{l}{\textbf{No S1}} & \multicolumn{1}{l}{\textbf{VulTrigger S1}} & \multicolumn{1}{l}{\textbf{Non-Fine-Tune/Powerful S1}} & \multicolumn{1}{l}{\textbf{No-Fine-Tune/Base S1}}\\
\hline
SEGV & \cellcolor{gray!20}\textbf{72.73\%} & 67.27\% & \cellcolor{gray!20}\textbf{72.73\%} & 69.09\% & 49.09\% & 21.82\% & 63.64\% & 62.50\% & 61.54\% & 56.25\% \\
global-buffer-overflow & \cellcolor{gray!20}\textbf{90.91\%} & 86.36\% & 81.82\% & 86.36\% & 81.82\% & 18.18\% & 63.64\% & 50.00\% & 75.00\% & 71.43\% \\
heap-buffer-overflow & \cellcolor{gray!20}\textbf{80.66\%} & 78.60\% & 74.49\% & 72.84\% & 60.08\% & 23.87\% & 62.14\% & 68.97\% & 71.09\% & 66.13\% \\
heap-use-after-free & 77.14\% & 77.14\% & \cellcolor{gray!20}\textbf{80.00\%} & 68.57\% & 60.00\% & 37.14\% & 74.29\% & 58.33\% & 75.00\% & 74.07\% \\
stack-buffer-overflow & \cellcolor{gray!20}\textbf{74.19\%} & 70.97\% & \cellcolor{gray!20}\textbf{74.19\%} & 70.97\% & 58.06\% & 12.90\% & 70.97\% & 44.44\% & 68.75\% & 50.00\% \\
stack-buffer-underflow & \cellcolor{gray!20}\textbf{100.00\%} & \cellcolor{gray!20}\textbf{100.00\%} & 66.67\% & 66.67\% & 66.67\% & 0.00\% & 66.67\% & 50.00\% & 0.00\% & 0.00\% \\
stack-use-after-return & 66.67\% & 66.67\% & 33.33\% & 66.67\% & 33.33\% & 0.00\% & 66.67\% & \cellcolor{gray!20}\textbf{100.00\%} & 0.00\% & \cellcolor{gray!20}\textbf{100.00\%} \\
undefined-behavior & 56.25\% & 56.25\% & 50.00\% & 56.25\% & 56.25\% & 6.25\% & 43.75\% & \cellcolor{gray!20}\textbf{100.00\%} & 28.57\% & 60.00\% \\
use-after-poison & 85.71\% & 71.43\% & 71.43\% & 71.43\% & 42.86\% & 28.57\% & 42.86\% & \cellcolor{gray!20}\textbf{100.00\%} & 33.33\% & \cellcolor{gray!20}\textbf{100.00\%} \\
use-of-uninitialized-value & \cellcolor{gray!20}\textbf{81.33\%} & 80.00\% & 77.33\% & 76.00\% & 61.33\% & 33.33\% & 70.67\% & 65.52\% & 65.31\% & 62.22\% \\
\hline
\textit{All} & \cellcolor{gray!35}\textit{\textbf{78.98\%}} & \textit{76.53\%} & \textit{74.29\%} & \textit{72.45\%} & \textit{59.39\%} & \textit{\underline{24.29\%}} & \textit{64.29\%} & \textit{65.38\%} & \textit{67.56\%} & \textit{64.59\%} \\
\hline
\end{tabular}}
\vspace{-4pt}
\label{tab:ablation}
\end{table*}

\subsection{RQ3: Ablation Studies} \label{sec:rq3}

To assess the contribution of each {\tech} component, we conduct ablation studies that isolate key design choices across data preparation, model specialization, and PoV generation.



\textbf{Impact of Neuro-Symbolic Design Choices.}
We first evaluate the impact of constraint processing and solver integration. 
As shown in Table~\ref{tab:ablation}, removing the SMT solver (``\emph{No SMT Solver}'') reduces the success rate from 78.98\% to 76.53\%, indicating that SMT fallback improves robustness on complex constraints; we disable it in subsequent ablations to isolate the remaining factors. Without constraint optimization/simplification (``\emph{No Simp}''), success further drops to 74.29\%, and removing both optimization and data augmentation (``\emph{No Simp\&Aug}'') reduces it to 72.45\%, showing that template-based augmentation and constraint optimization jointly contribute to effective PoV generation.



\textbf{Impact of Constraint-Reasoning Fine-Tuning.}
We next evaluate the importance of fine-tuning for constraint reasoning by replacing the fine-tuned CEM and CSM with a base model (``\emph{No-Fine-Tune/Base S3}'', Llama-3.1-8B) and a powerful commercial LLM (``\emph{Non-Fine-Tune/Powerful S3}'', Gemini-2.5-Flash). We use Gemini-2.5-Flash rather than Gemini-2.5-Pro here because the ablation queries the model extensively---4.57M tokens per sample on average (\$1.37/sample with Flash vs.\ \$5.71 with Pro)---whereas the RQ2 baseline issues a single direct prompt and can afford Pro. As shown in Table~\ref{tab:ablation}, success drops substantially to 59.39\% with Gemini and 24.29\% with Llama3, showing that general-purpose LLMs, even powerful ones, may not match task-specific fine-tuning.

\textbf{Direct Accuracy Evaluation of CEM and CSM.}
To complement the end-to-end results, we directly evaluate CEM and CSM accuracy on the 52-sample RQ2 subset, scoring the top 50 execution paths per sample. For CEM, two authors blindly annotated each constraint set for \emph{syntax validity} and \emph{semantics correctness} (inter-annotator Cohen's $\kappa = 0.81$), and we report precision and recall. As shown in Table~\ref{tab:direct-accuracy}, the fine-tuned CEM attains the highest recall---the most relevant metric, since one correct constraint set per sample suffices for a PoV---matching or slightly exceeding Claude-Sonnet-4 at no per-sample API cost (a \$107 difference in API spend across the 52 samples), while the base Llama lags on every metric. For CSM, we feed each candidate solver the same CEM-extracted constraints and report PoV success. As also shown in Table~\ref{tab:direct-accuracy}, the fine-tuned CSM achieves the highest success at no API cost; the \emph{Z3 only} baseline---which solves the same constraints without CSM---falls well below CSM, isolating its contribution and explaining why it far exceeds whole-program KLEE despite both relying on Z3.

\textbf{Impact of Manifestation Localization.}
Finally, we evaluate the importance of Step~1. Removing manifestation localization (``\emph{No S1}'') and relying only on paths to the patch location reduces the success rate from 78.98\% to 64.29\%, indicating that patch locations alone are insufficient for effective exploration. Comparing the fine-tuned MLM against alternatives---VulTrigger~\cite{li2024effectiveness} and general-purpose LLMs (Gemini-2.5-Flash and Llama-3.2-3B)---the MLM achieves the best localization performance (Table~\ref{tab:loc-acc}), and the ``\emph{Non-Fine-Tune/Powerful S1}'' and ``\emph{No-Fine-Tune/Base S1}'' rows in Table~\ref{tab:ablation} show that weaker localization yields lower end-to-end PoV generation success. Notably, {\tech}'s end-to-end success rate (78.98\%) closely matches its Step~1 recall (79.18\%): as long as the predicted location lies on the correct vulnerable control-flow path, Steps~2 and 3 tolerate small localization errors, though high-recall localization remains essential since incorrect anchors impede effective path exploration.

\begin{table}[t]
 \centering
 \vspace{-12pt}
 \caption{Average per-sample time (minutes) and LLM cost (\$) of \textsc{PoVGen} vs.\ baselines}
 \vspace{-8pt}
 \label{tab:efficiency}
\scalebox{0.75}{
\begin{tabular}{lrr|lrr|lrr}
\hline
\textbf{\textsc{PoVGen}} & \textbf{Time} & \textbf{\$} & \textbf{Fuzz/SymEx} & \textbf{Time} & \textbf{\$} & \textbf{LLM} & \textbf{Time} & \textbf{\$} \\
\hline
\cellcolor{gray!15}\emph{\textbf{Total}} & \cellcolor{gray!15}\emph{\textbf{98.21}} & \cellcolor{gray!15}\emph{\textbf{0.04}} & libFuzzer & 260.49 & 0 & Gemini-2.5-Pro & \cellcolor{green!10}8.23 & 1.72 \\
Step 1 & 0.61 & 0.04 & AFL++ & 729.38 & 0 & OpenAI-o3 & \cellcolor{green!10}\textbf{5.64} & 1.37 \\
Step 2 & 75.71 & 0 & AFLGo & 625.17 & 0 & Claude-4 & \cellcolor{green!10}\textbf{4.32} & \cellcolor{red!10}\underline{2.06} \\
Step 3 & 21.89 & 0 & KLEE & \underline{1122.23} & 0 & & & \\
\hline
\end{tabular}
}
\vspace{-10pt}
\end{table}

\vspace{-2pt}
\subsection{RQ4: Efficiency}


Table~\ref{tab:efficiency} reports the efficiency of {\tech} and the baselines in terms of runtime and monetary cost.

\begin{figure}[t]
\centering
\vspace{-0pt}
	\includegraphics[width=1.0\linewidth]{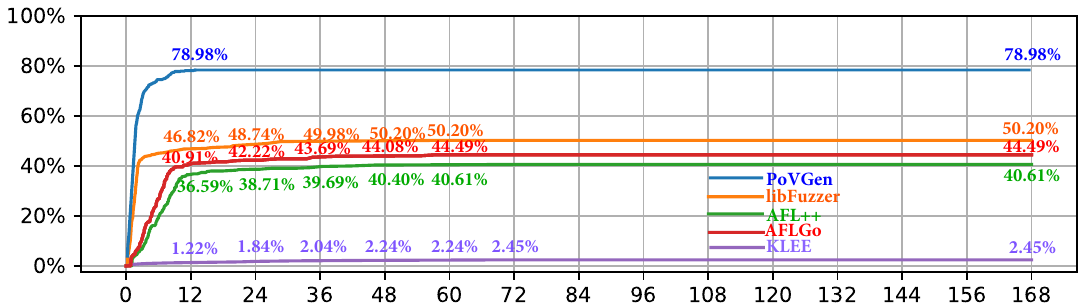}
        \vspace{-20pt}
        \caption{Progression of success rate ($y$ axis) achieved at increasing time cost in terms of hours spent ($x$ axis). }
	\label{fig:fuzzing-time}
        \vspace{-0pt}
\end{figure}



{\tech} requires an average of 98.21 minutes to reproduce a vulnerability, most of it in Step~2 (\textit{Path-Sensitive Reachability Exploration}, 75.71 min) and Step~3 (\textit{Neuro-Symbolic PoV Generation}, 21.89 min); only Step~1 incurs an API cost, averaging \$0.04 per sample for GLM-5.2, while the remaining steps run locally at no per-sample cost. In contrast, fuzzing is far slower---LibFuzzer, AFL++, and AFLGo plateau after roughly 48 hours (Figure~\ref{fig:fuzzing-time}) and take 260--729 minutes per successful sample (Table~\ref{tab:efficiency})---and KLEE is least efficient at 1{,}122 minutes for the small fraction it reproduces, owing to path explosion in whole-program symbolic execution.


\subsection{RQ5: Real-World CVE Reproduction}\label{sec:rq5}

\begin{table}[tp]
 \centering
 \vspace{-12pt}
     \caption{Effectiveness of {\tech} on real-world CVEs without public PoVs (\#samples in parentheses)}
\vspace{-7pt}
\scalebox{0.72}{
\begin{tabular}{lr|lr|lr}
\hline
\textbf{CWE ID} & \multicolumn{1}{l|}{\textbf{\%Success}} & \textbf{CWE ID} & \multicolumn{1}{l|}{\textbf{\%Success}} & \textbf{CWE ID} & \multicolumn{1}{l}{\textbf{\%Success}} \\
\hline
CWE-119 (39) & \underline{66.67\%} & CWE-190 (1) & \cellcolor{gray!20}\textbf{100.00\%} & CWE-415 (9) & 77.78\% \\
CWE-120 (9) & 88.89\% & CWE-193 (1) & \cellcolor{gray!20}\textbf{100.00\%} & CWE-416 (23) & 73.91\% \\
CWE-122 (2) & \cellcolor{gray!20}\textbf{100.00\%} & CWE-20 (22) & 81.82\% & CWE-476 (25) & 72.00\% \\
CWE-125 (69) & 78.26\% & CWE-399 (1) & \cellcolor{gray!20}\textbf{100.00\%} & CWE-787 (47) & 68.09\% \\
\cline{5-6}CWE-1284 (1) & \cellcolor{gray!20}\textbf{100.00\%} & CWE-401 (1) & \cellcolor{gray!20}\textbf{100.00\%} & \textit{All (250)} & \cellcolor{gray!30}\textit{\textbf{74.80\%}} \\
\hline
\end{tabular}
}
\vspace{4pt}
\label{tab:cve-results}
\end{table}


We evaluate {\tech} on 250 real-world CVEs for which no public PoVs are available. As shown in Table~\ref{tab:cve-results}, {\tech} successfully generates valid PoVs for 74.80\% of these vulnerabilities, demonstrating its ability to reproduce vulnerabilities beyond curated benchmarks, with consistently high success rates across CWE categories---perfect performance on several classes (e.g., CWE-122, CWE-190, CWE-401) and strong results on common types such as CWE-125 and CWE-415. The variation across CWEs reflects differences in constraint complexity and input structure: vulnerabilities with localized, well-defined triggers are easier to reproduce than those needing complex inputs or deeper execution contexts.

\section{Discussion}
\vspace{-0pt}
\subsection{Why \textsc{PoVGen} Works}
\vspace{-0pt}
\label{sec:why-povgen-works}

\textsc{PoVGen}'s effectiveness stems from decomposing PoV generation into three sub-problems, each addressed by a mechanism that targets one of the three challenges in \S\ref{sec:bkgmot}. (1)~Anchoring exploration to the manifestation point converts an otherwise intractable whole-program search into a focused reachability problem over a small set of relevant paths (\textit{Challenge 1}). (2)~Reasoning over these focused paths with the fine-tuned CEM and CSM, backed by an SMT solver, derives and satisfies path constraints without full-program symbolic execution (\textit{Challenge 2}). (3)~Fine-tuning on symbolic-execution-derived data bridges the gap between general-purpose LLM reasoning and these formal sub-tasks, while local deployment avoids per-sample API cost (\textit{Challenge 3}).

In contrast, applying general-purpose LLMs directly is insufficient: without semantic focusing the model cannot localize the vulnerable region, and even given a correct path it struggles to produce valid constraints and inputs without task-specific fine-tuning (\S\ref{sec:rq2}, \S\ref{sec:rq3}). Both search-space narrowing and fine-tuning are thus necessary for effective PoV generation.

\begin{table}[tp]
 \vspace{-0pt}
 \centering
     \caption{Flawed patches found by {\tech}} 
     \vspace{-8pt}
\scalebox{0.8}{
\begin{tabular}{llll}
\hline
CVE   & Project & CWE   & Patch Issue \\
\hline
\href{https://nvd.nist.gov/vuln/detail/CVE-2020-15471}{CVE-2020-15471} & \href{https://github.com/ntop/nDPI}{nDPI} & CWE-125 & Incomplete \\
\href{https://nvd.nist.gov/vuln/detail/CVE-2020-15473}{CVE-2020-15473} & \href{https://github.com/ntop/nDPI}{nDPI} & CWE-125 & Incorrect \\
\href{https://nvd.nist.gov/vuln/detail/CVE-2020-15475}{CVE-2020-15475} & \href{https://github.com/ntop/nDPI}{nDPI} & CWE-416 & Incomplete \\
\href{https://nvd.nist.gov/vuln/detail/CVE-2020-26570}{CVE-2020-26570} & \href{https://github.com/OpenSC/OpenSC}{OpenSC} & CWE-787 & Incorrect \\
\href{https://nvd.nist.gov/vuln/detail/CVE-2022-23537}{CVE-2022-23537} & \href{https://github.com/pjsip/pjproject}{PJSIP} & CWE-125 & Incomplete \\
\href{https://nvd.nist.gov/vuln/detail/CVE-2023-27599}{CVE-2023-27599} & \href{https://github.com/OpenSIPS}{OpenSIPS} & CWE-20 & Incomplete \\
\hline
\end{tabular}%
}
\vspace{-0pt}
\label{tab:bad-patches}
\end{table}

\vspace{-2pt}
\subsection{Additional Usefulness of {\tech}}\label{sec:useful}
\vspace{-2pt}



Generated PoVs enable not only reproduction of known vulnerabilities but also validation of patches: applying each CVE's PoV to its patched version reveals six flawed patches (Table~\ref{tab:bad-patches})---four incomplete (the PoV still triggers the original crash) and two incorrect (the PoV triggers a different crash). They can also surface new vulnerabilities: applying those from the ARVO dataset to more recent versions of the same projects revealed five previously unreported vulnerabilities (Table~\ref{tab:zero-days}), which we reported via responsible disclosure. Beyond these, the PoVs provide high-quality ground-truth data for training analysis tools and concrete debugging artifacts, and by linking each PoV to its triggering execution path, {\tech} aids root-cause analysis and patch development.

\vspace{-2pt}
\subsection{Comparison with Closed-Model LLM-Assisted Systems}
\vspace{-2pt}
PoCGen~\cite{simsek2025pocgen}, ConcoLLMic~\cite{luo2026agentic}, and CottonTail~\cite{tu2026cottontail} represent important LLM-assisted design points. PoCGen uses a closed-weight, frontier LLM in a synthesis-and-validate loop for PoC construction, while ConcoLLMic and CottonTail also use closed-weight, frontier LLMs to assist dynamic/concolic exploration, where concrete executions or generated seeds guide subsequent input generation. {\tech} studies a \textit{different point in the design space}: vulnerability-specific PoV generation without closed-model dependence, using pinned open-weight model artifacts, symbolic path guidance, and SMT-backed constraint reasoning. Our goal is not to position {\tech} as a replacement for these systems or to attempt to surpass them, but to study a complementary question: 
\textit{how effective PoV generation can be under a more cost-conscious and controllable design}. 
A head-to-head comparison would primarily evaluate a different question---how proprietary LLM-assisted systems perform with evolving frontier models and agentic or concolic scaffolds---whereas this paper focuses on what can be achieved with pinned open-weight models and symbolic reasoning. 
We thus view these approaches as complementary rather than substitutable baselines.

\begin{table}[tp]
 \vspace{-4pt}
 \centering
     \caption{Previously unreported vulnerabilities found by {\tech}} 
     \vspace{-5pt}
\scalebox{0.8}{
\begin{tabular}{lll}
\hline
Project & Version & Vulnerability Type \\
\hline
\href{https://github.com/vstakhov/libucl}{libucl} & 0.9.1 & Heap-buffer-overflow \\
\href{https://github.com/vstakhov/libucl}{libucl} & 0.9.1 & Memory leak \\
\href{https://github.com/radareorg/radare2}{radare2} & 5.9.2 & Memory leak \\
\href{https://github.com/tbeu/matio}{matio} & 1.5.27 & Memory leak \\
\href{https://github.com/GStreamer/gstreamer}{gstreamer} & 1.24.3 & Stack-buffer-overflow \\
\hline
\end{tabular}
}
\vspace{8pt}
\label{tab:zero-days}
\end{table}

\subsection{Threats to Validity}
\vspace{-2pt}
\textbf{Internal validity.} Because ARVO and disclosed CVEs are public, they may appear in the pretraining corpora of large language models. To mitigate this, {\tech} relies on \emph{fine-tuned, open-weight} models, with the only general-purpose model (GLM-5.2, for slicing) being open-weight and its suggestions verified against the source tree; the ablations (\S\ref{sec:rq3}) further indicate that the gains stem from semantic focusing and fine-tuning rather than memorization, since general-purpose substitutes---which would benefit equally from any memorized data---perform substantially worse. Because every pipeline model is open-weight, {\tech} is also 
reproducible and 
controllable 
without closed commercial APIs, and we release the framework, fine-tuning data, and evaluation pipeline.

\textbf{External validity.} Our evaluation targets C/C++ memory-safety vulnerabilities, the most prevalent class with high-quality benchmarks; extending to other types and languages is left to future empirical validation.

\section{Related Work}

\textbf{Symbolic execution and fuzzing.} Classic Automatic Exploit Generation combines symbolic execution and SMT solving (KLEE~\cite{cadar2008klee}, SAGE~\cite{godefroid2012sage}), with hybrid and patch-based variants adding fuzzing or binary diffing to reach deeper code (Driller~\cite{stephens2017driller}, AFL~\cite{zalewski2017american}, AFL++~\cite{fioraldi2020afl++}, patch-based AEG~\cite{brumley2008aeg}), and directed  fuzzers steer exploration toward target sites (AFLGo~\cite{bohme2017directed}). These approaches are dynamic and rely on seed inputs or exhaustive path search, whereas {\tech} statically focuses on vulnerability-relevant paths and needs no seeds.

\textbf{LLM-based approaches.} Recent work applies learning and LLMs to bug finding and exploitation: imitation learning for contract fuzzing~\cite{he2019learning}; agentic concolic execution and LLM-driven constraint solving (ConcoLLMic~\cite{luo2026agentic}, LangSym~\cite{xu2024symbolic}, AutoBug~\cite{li2025large}, SymGPT~\cite{xia2025symgpt}); and direct LLM prompting for PoV generation, vulnerability tracing, and testing (PoCGen~\cite{simsek2025pocgen}, FaultLine~\cite{nitin2025faultline}, PwnGPT~\cite{peng2025pwngpt}, PentestGPT~\cite{deng2024pentestgpt}). {\tech} instead synthesizes vulnerability-specific PoVs using fine-tuned open-weight models and symbolic reasoning.

\textbf{Vulnerability-specific generators.} Other systems target narrow classes or environments---PHP object injection (Fugio~\cite{park2022fugio}), Node.js packages (Explode.js~\cite{marques2025explode}), web applications (NAVEX~\cite{alhuzali2018navex}), and kernel or interpreter heap overflows (KOOBE~\cite{chen2020koobe}, Gollum~\cite{heelan2019gollum}, AAHEG~\cite{wang2023aaheg})---and are specialized to particular languages or runtimes. {\tech} instead offers a general pipeline for C/C++ memory-safety vulnerabilities, decoupled from class-specific heuristics.




\vspace{-0pt}
\section{Conclusion}
\vspace{-0pt}
We presented {\tech}, a neuro-symbolic framework for generating PoVs 
where no triggering/seed inputs are available. By combining manifestation localization, path-sensitive reachability exploration, and LLM-guided constraint reasoning, {\tech} 
achieves 
stronger effectiveness than fuzzing and symbolic-execution baselines, together with lower model-inference cost than direct frontier-LLM prompting in controlled comparisons.
Its generated PoVs further revealed six flawed patches and five previously unreported vulnerabilities, demonstrating practical value beyond reproduction. Artifacts are available at \url{https://figshare.com/s/898d1712d8c814820262}.

\vspace{-5pt}

\bibliography{paper-abbr2}

@inproceedings{serebryany2016continuous,
  title={Continuous fuzzing with libfuzzer and addresssanitizer},
  author={Serebryany, Kosta},
  booktitle={SecDev},
  year={2016}
}

@inproceedings{fioraldi2020afl++,
  title={{AFL++}: Combining incremental steps of fuzzing research},
  author={Fioraldi, Andrea and Maier, Dominik and Ei{\ss}feldt, Heiko and Heuse, Marc},
  booktitle={WOOT},
  year={2020}
}

@inproceedings{bohme2017directed,
  title={Directed greybox fuzzing},
  author={B{\"o}hme, Marcel and Pham, Van-Thuan and Nguyen, Manh-Dung and Roychoudhury, Abhik},
  booktitle={CCS},
  year={2017}
}

@inproceedings{pearce2023examining,
  title={Examining zero-shot vulnerability repair with large language models},
  author={Pearce, Hammond and Tan, Benjamin and Ahmad, Baleegh and Karri, Ramesh and Dolan-Gavitt, Brendan},
  booktitle={S\&P},
  year={2023},
}

@article{shahandashti2024program,
  title={Program slicing in the era of large language models},
  author={Shahandashti, Kimya Khakzad and Mohajer, Mohammad Mahdi and Belle, Alvine Boaye and Wang, Song and Hemmati, Hadi},
  journal={arXiv preprint arXiv:2409.12369},
  year={2024}
}

@inproceedings{park2022fugio,
  title={{FUGIO}: Automatic exploit generation for {PHP} object injection vulnerabilities},
  author={Park, Sunnyeo and Kim, Daejun and Jana, Suman and Son, Sooel},
  booktitle={USENIX Security},
  year={2022}
}

@inproceedings{he2019learning,
  title={Learning to fuzz from symbolic execution with application to smart contracts},
  author={He, Jingxuan and Balunovi{\'c}, Mislav and Ambroladze, Nodar and Tsankov, Petar and Vechev, Martin},
  booktitle={CCS},
  year={2019}
}

@article{wang2023aaheg,
  title={{AAHEG}: Automatic Advanced Heap Exploit Generation Based on Abstract Syntax Tree},
  author={Wang, Yu and Zhang, Yipeng and Li, Zhoujun},
  journal={Symmetry},
  year={2023},
}

@article{simsek2025pocgen,
  title={{PoCGen}: Generating Proof-of-Concept Exploits for Vulnerabilities in Npm Packages},
  author={Simsek, Deniz and Eghbali, Aryaz and Pradel, Michael},
  journal={arXiv preprint arXiv:2506.04962},
  year={2025}
}

@article{nitin2025faultline,
  title={FaultLine: Automated Proof-of-Vulnerability Generation Using LLM Agents},
  author={Nitin, Vikram and Ray, Baishakhi and Moghaddam, Roshanak Zilouchian},
  journal={arXiv preprint arXiv:2507.15241},
  year={2025}
}

@inproceedings{luo2026agentic,
  title={Agentic Concolic Execution},
  author={Luo, Zhengxiong and Zhao, Huan and Wolff, Dylan and Cadar, Cristian and Roychoudhury, Abhik},
  booktitle={Proceedings of the IEEE Symposium on Security and Privacy (S\&P)},
  pages={1--19},
  year={2026}
}

@misc{glm52,
  author       = {{Z.ai}},
  title        = {{GLM-5.2}: An Open-Weight Large Language Model},
  year         = {2026},
  howpublished = {\url{https://huggingface.co/zai-org/GLM-5.2}},
}

@article{li2025large,
  title={Large language model powered symbolic execution},
  author={Li, Yihe and Meng, Ruijie and Duck, Gregory J},
  journal={OOPSLA},
  year={2025},
}

@inproceedings{tu2026cottontail,
  title={Cottontail: Large Language Model-Driven Concolic Execution for Highly Structured Test Input Generation},
  author={Tu, Haoxin and Lee, S and Li, Y and Chen, P and Jiang, L and B{\"o}hme, Marcel},
  booktitle={Proceedings of the IEEE Symposium on Security and Privacy (S\&P). IEEE},
  pages={1--19},
  year={2026}
}

@inproceedings{xu2024symbolic,
  title={Symbolic Execution with Test Cases Generated by Large Language Models},
  author={Xu, Jiahe and Xu, Jingwei and Chen, Taolue and Ma, Xiaoxing},
  booktitle={2024 IEEE 24th International Conference on Software Quality, Reliability and Security (QRS)},
  pages={228--237},
  year={2024},
  organization={IEEE}
}

@article{xia2025symgpt,
  title={{SymGPT}: Auditing Smart Contracts via Combining Symbolic Execution with Large Language Models},
  author={Xia, Shihao and He, Mengting and Shao, Shuai and Yu, Tingting and Zhang, Yiying and Song, Linhai},
  journal={arXiv preprint arXiv:2502.07644},
  year={2025}
}

@inproceedings{li2024effectiveness,
  title={On the Effectiveness of Function-Level Vulnerability Detectors for Inter-Procedural Vulnerabilities},
  author={Li, Zhen and Wang, Ning and Zou, Deqing and Li, Yating and Zhang, Ruqian and Xu, Shouhuai and Zhang, Chao and Jin, Hai},
  booktitle={ICSE},
  year={2024}
}

@article{zhou2024largeb,
  title={Large Language Model for Vulnerability Detection and Repair: Literature Review and Roadmap},
  author={Zhou, Xin and Cao, Sicong and Sun, Xiaobing and Lo, David},
  journal={arXiv preprint arXiv:2404.02525},
  year={2024}
}

@inproceedings{li2017large,
  title={A large-scale empirical study of security patches},
  author={Li, Frank and Paxson, Vern},
  booktitle={CCS},
  year={2017}
}

@inproceedings{wu2025veribin,
  title={VeriBin: Adaptive Verification of Patches at the Binary Level.},
  author={Wu, Hongwei and Wu, Jianliang and Wu, Ruoyu and Sharma, Ayushi and Machiry, Aravind and Bianchi, Antonio},
  booktitle={NDSS},
  year={2025}
}

@misc{kikta2024patch,
    Author={{Jason Kikta}},
    Howpublished={\url{https://www.automox.com/blog/why-automated-patching-is-critical}},
    Title={The Shrinking Window for Vulnerability Exploitation: Why Automated Patching is Critical},
    Year = {2025}
}

@article{serebryany2017oss,
  title={{OSS-Fuzz}-Google's continuous fuzzing service for open source software},
  author={Serebryany, Kostya},
  year={2017}
}

@article{li2021sysevr,
  title={{SySeVR}: A framework for using deep learning to detect software vulnerabilities},
  author={Li, Zhen and Zou, Deqing and Xu, Shouhuai and Jin, Hai and Zhu, Yawei and Chen, Zhaoxuan},
  journal={TDSC},
  year={2021}
}

@misc{joern,
    Author={Fabian Yamaguchi},
    Howpublished={\url{https://joern.readthedocs.io/en/latest/installation.html}},
    Title={A platform for robust analysis of C/C++ code},
    Year = {2022}
}

@misc{nvd,
    Author={{National Institute of Standards and Technology (NIST)}},
    Howpublished={\url{https://nvd.nist.gov}},
    Title={{National Vulnerability Database (NVD)}},
    Year = {2025}
}

@misc{exploitdb,
    Author={{OffSec}},
    Howpublished={\url{https://www.exploit-db.com/}},
    Title={{The Exploit Database}},
    Year = {2025}
}

@inproceedings{brumley2008aeg,
  title={Automatic patch-based exploit generation is possible: Techniques and implications},
  author={Brumley, David and Poosankam, Pongsin and Song, Dawn and Zheng, Jiang},
  booktitle={S\&P},
  year={2008},
}

@misc{vulconsequence231,
    Author={{Ericsson}},
    Howpublished = {\url{https://www.ericsson.com/en/security/vulnerability-management}},
    Title={Software vulnerability: Impact \& ways to avoid it},
    year = {2023}
}

@misc{cvedashboard23,
    Author={{NIST}},
    Howpublished = {\url{https://nvd.nist.gov/general/nvd-dashboard}},
    Title={National Vulnerability Database ({NVD}) Dashboard},
    year = {2023}
}

@misc{wllvm,
    Author={{Tristan Ravitch}},
    Howpublished = {\url{https://github.com/travitch/whole-program-llvm}},
    Title={A wrapper script to build whole-program LLVM bitcode files},
    year = {2021}
}

@misc{openai-o3,
    Author={{OpenAI}},
    Howpublished = {\url{https://openai.com/index/introducing-o3-and-o4-mini/}},
    Title={Introducing OpenAI o3 and o4-mini},
    year = {2025}
}

@inproceedings{peng2025pwngpt,
  title = "{P}wn{GPT}: Automatic Exploit Generation Based on Large Language Models",
  author = "Wanzong Peng and Lin Ye and Xuetao Du and Hongli Zhang and Dongyang Zhan and Yunting Zhang and Yicheng Guo and Chen Zhang",
  booktitle = "ACL",
  year = "2025",
}

@inproceedings{deng2024pentestgpt,
  title={{PentestGPT}: Evaluating and harnessing large language models for automated penetration testing},
  author={Deng, Gelei and Liu, Yi and Mayoral-Vilches, V{\'\i}ctor and Liu, Peng and Li, Yuekang and Xu, Yuan and Zhang, Tianwei and Liu, Yang and Pinzger, Martin and Rass, Stefan},
  booktitle={USENIX Security},
  year={2024}
}

@article{godefroid2012sage,
author = {Godefroid, Patrice and Levin, Michael Y. and Molnar, David},
title = {{SAGE}: whitebox fuzzing for security testing},
year = {2012},
journal = {Communication of ACM},
}

@misc{zalewski2017american,
  title={American fuzzy lop ({AFL}) fuzzer},
  author={Zalewski, Michal},
  Howpublished={\url{https://lcamtuf.coredump.cx/afl}},
  year={2017}
}

@inproceedings{alhuzali2018navex,
  title={{NAVEX}: Precise and scalable exploit generation for dynamic web applications},
  author={Alhuzali, Abeer and Gjomemo, Rigel and Eshete, Birhanu and Venkatakrishnan, VN},
  booktitle={USENIX Security},
  year={2018}
}

@article{marques2025explode,
  title={Automated Exploit Generation for {Node.js} Packages},
  author={Marques, Filipe and Ferreira, Mafalda and Nascimento, Andr{\'e} and Coimbra, Miguel E and Santos, Nuno and Jia, Limin and Fragoso Santos, Jos{\'e}},
  journal={PLDI},
  year={2025},
}

@inproceedings{chen2020koobe,
  title={{KOOBE}: Towards facilitating exploit generation of kernel Out-Of-Bounds write vulnerabilities},
  author={Chen, Weiteng and Zou, Xiaochen and Li, Guoren and Qian, Zhiyun},
  booktitle={USENIX security},
  year={2020}
}

@inproceedings{heelan2019gollum,
  title={Gollum: Modular and greybox exploit generation for heap overflows in interpreters},
  author={Heelan, Sean and Melham, Tom and Kroening, Daniel},
  booktitle={CCS},
  year={2019}
}

@article{wang2025pbaeg,
  title={{PBAEG}: combine-vulnerabilities AEG to defeat protection mechanisms},
  author={Wang, Yu and Li, Zhoujun and Zhang, Yipeng},
  journal={Cybersecurity},
  year={2025},
}

@inproceedings{stephens2017driller,
  title = "{Driller}: Augmenting Fuzzing Through Selective Symbolic Execution",
  author = "Nick Stephens and John Grosen and Christopher Salls and Andrew Dutcher and Ruoyu Wang and Jacopo Corbetta and Yan Shoshitaishvili and Christopher Kruegel and Giovanni Vigna",
  booktitle = "NDSS",
  year = "2017"
}

@book{black2018juliet,
  title={Juliet 1.3 test suite: Changes from 1.2},
  author={Black, Paul E},
  year={2018},
  publisher={US Department of Commerce, National Institute of Standards and Technology}
}

@inproceedings{beyer2021software,
  title={Software Verification: 10th Comparative Evaluation ({SV-COMP} 2021)},
  author={Beyer, Dirk},
  booktitle={TACAS},
  year={2021},
}

@article{mei2024arvo,
  title={{ARVO}: Atlas of reproducible vulnerabilities for open source software},
  author={Mei, Xiang and Singaria, Pulkit Singh and Del Castillo, Jordi and Xi, Haoran and Bao, Tiffany and Wang, Ruoyu and Shoshitaishvili, Yan and Doup{\'e}, Adam and Pearce, Hammond and Dolan-Gavitt, Brendan and others},
  journal={arXiv preprint arXiv:2408.02153},
  year={2024}
}

@misc{han2024unsloth,
  title={Unsloth},
  author={Han, Daniel and Han, Michael},
  HowPublished={\url{https://github. com/unslothai/unsloth}},
  year={2024}
}

@inproceedings{de2008z3,
  title={Z3: An efficient SMT solver},
  author={De Moura, Leonardo and Bj{\o}rner, Nikolaj},
  booktitle={TACAS},
  year={2008},
}

@inproceedings{cadar2008klee,
  title={{KLEE}: unassisted and automatic generation of high-coverage tests for complex systems programs.},
  author={Cadar, Cristian and Dunbar, Daniel and Engler, Dawson R and others},
  booktitle={OSDI},
  year={2008}
}

@inproceedings{sui2016svf,
  title={{SVF}: interprocedural static value-flow analysis in LLVM},
  author={Sui, Yulei and Xue, Jingling},
  booktitle={CC},
  year={2016}
}

@inproceedings{wu2023effective,
  title={How effective are neural networks for fixing security vulnerabilities},
  author={Wu, Yi and Jiang, Nan and Pham, Hung Viet and Lutellier, Thibaud and Davis, Jordan and Tan, Lin and Babkin, Petr and Shah, Sameena},
  booktitle={ISSTA},
  year={2023}
}

@article{horwitz1990interprocedural,
  title={Interprocedural slicing using dependence graphs},
  author={Horwitz, Susan and Reps, Thomas and Binkley, David},
  journal={TOPLAS},
  year={1990},
}

@book{cochran1977sampling,
  author    = {Cochran, William G.},
  title     = {Sampling Techniques},
  edition   = {3rd},
  publisher = {Wiley},
  year      = {1977}
}
\bibliographystyle{IEEEtran}

\end{document}